\documentclass[twocolumn]{aastex631}

\usepackage{bm}
\usepackage{multirow}
\usepackage{amsmath}
\usepackage{color}
\usepackage{subfigure}

\newcommand{\hi}{\textsc{Hi}}
\newcommand{\hh}{\ensuremath{\rm H_2}}
\newcommand{\hihh}{\textsc{Hi}/\ensuremath{\rm {H_2}}}
\newcommand{\hii}{\textsc{Hii}}

\newcommand{\mhalo}{\ensuremath{\textrm{M}_{\textrm{h}}}}
\newcommand{\mhi}{\ensuremath{\textrm{M}_{\textsc{Hi}}}}
\newcommand{\msun}{\>{\rm M_{\odot}}}

\newcommand{\mstar}{\ensuremath{\textrm{M}_{\ast}}}
\newcommand{\logmstar}{\ensuremath{\log\mstar}}

\newcommand{\mbhdot}{\ensuremath{\dot{\textrm{M}}_\textrm{BH}}}
\newcommand{\mbh}{\ensuremath{\textrm{M}}_\textrm{BH}}
\newcommand{\hmpc}{\>{h^{-1}{\rm Mpc}}}
\newcommand{\hmsun}{\ensuremath{\>{h^{-1}{\rm M_{\odot}}}}}
\newcommand{\lgfhi}{\ensuremath \log {\rm M_{\hi}/M_*}}
\newcommand{\fhi}{\ensuremath{\rm M_{\hi}/M_*}}
\newcommand{\logssfr}{\ensuremath{\rm \log sSFR}}
\newcommand{\logmbh}{\ensuremath{\rm \log M_{BH}}}
\newcommand{\logmbhdot}{\ensuremath{\rm \log \dot{M}_{BH}}}
\newcommand{\logoh}{\ensuremath{\rm \log (O/H) + 12}}

\shorttitle{\hi\ in IllustrisTNG, EAGLE and xGASS}
\shortauthors{Li, Li \& Mo}

\begin{document}

\title{What drives the H{\sc i} content of central galaxies - A comparison between hydrodynamic simulations and observations using Random Forest}

\correspondingauthor{Xiao Li}
\email{xli27938@gmail.com}
\correspondingauthor{Cheng Li}
\email{cli2015@tsinghua.edu.cn}

\author[0000-0002-2884-9781]{Xiao Li}
\affiliation{Department of Astronomy, Tsinghua University, Beijing 100084, China}

\author[0000-0002-8711-8970]{Cheng Li}
\affiliation{Department of Astronomy, Tsinghua University, Beijing 100084, China}

\author[0000-0001-5356-2419]{H. J. Mo}
\affiliation{Department of Astronomy, University of Massachusetts Amherst, MA 01003, USA}

\begin{abstract}



We investigate the driving mechanisms for the \hi\ gas content in star-forming central galaxies at low redshift, by examining the \hi-to-stelalr mass ratio ($M_{\rm HI}/M_\ast$) in both the state-of-the-art hydrodynamic simulations, IllustrisTNG (TNG) and EAGLE, and the xGASS sample. We quantify the correlations of  $M_{\rm HI}/M_\ast$ with a variety of galaxy properties using the random forest regression technique, and we make comparisons between the two simulations, as well as between the simulations and xGASS. Gas-phase metallicity is found to be most important in both simulations, but is ranked mildly for xGASS, suggesting that metals and gas driven by feedback effects in real galaxies is not as tightly coupled as in the simulations. Beyond that, the accretion rate of supermassive black holes is the most important feature in TNG, while specific star formation rate is the top ranked in EAGLE. This result can be understood from the fact that the \hi\ gas is regulated mainly by thermal-mode AGN feedback in TNG and by stellar feedback in EAGLE.  Although neither simulation can fully reproduce the feature importance obtained for real galaxies in the xGASS, EAGLE performs better than TNG in the sense that the observationally top-ranked property, $u-r$, is also highly ranked in EAGLE. This result implies that stellar feedback plays a more dominant role than AGN feedback in driving the \hi\ gas content of low-redshift galaxies.   
\end{abstract}

\section{Introduction} \label{sec:intro}
In current models, galaxies form at the center of dark matter halos through cooling and condensation of gas \citep{White&Rees1978, MoBoschWhite2010}.  
Theoretically, gas-related processes in/around galaxies have been extensively studied in the past decade with the help of cosmological hydrodynamic simulations, such as Horizon-AGN \citep{Horizon-AGN}, Magneticum\footnote{http://www.magneticum.org/simulations.html}, EAGLE \citep{Crain_2015,Schaye_2015}, FIRE \citep{GIZMO,FIRE}, Illustris \citep{Vogelsberger2014Nature, Vogelsberger2014MNRAS, Sijacki2015, Genel2014}, IllustrisTNG \citep[][]{Springel_2018,Naiman_2018,Marinacci_2018,Nelson_2018,Pillepich_2018}, and SIMBA \citep{SIMBA}. In particular, many efforts have been made to investigate atomic hydrogen (\hi), the dominant component of cold gas. These simulations are usually able to reproduce the abundance of galaxies as a function of the \hi\ mass and the \hi\ size-mass relation in the local Universe \citep[e.g.][]{Bahe2016,Crain2017,Diemer2019,Ma2022}. However, the simulations have also shown that the \hi\ content of a galaxy can be regulated by a variety of processes, including consumption by star formation \citep{Springel_2003,Schaye_2008,FIRE}, heating or outflows  by stellar feedback \citep{Pillepich_2018,Dalla_2012,FIRE} and active galactic nucleus (AGN) feedback \citep{Booth2009,Horizon-AGN,Weinberger2017}, and tidal or ram-pressure stripping by surrounding hot gas and companion galaxies \citep{Marasco2016,Stevens2019,Watts2020}. Different simulations use different mechanisms to reproduce the observations of \hi\ gas, and it is unclear which mechanisms drive the \hi\ mass and distribution in real galaxies. For instance, using EMP-\textit{pathfinder} \citep{Reina-Campos2022} and \textsc{FIRE}box \citep{Feldmann2023}, \cite{Gensior2024} showed that different subgrid physics can result in consistent galaxy-wide \hi\ properties, but the small-scale properties are very different, with the rotational asymmetry of the \hi\ discs has the strongest dependence on the physical processes implemented.


On the observational side, large \hi\ surveys accomplished in the past two decades, such as \hi\ Parkes All-Sky Survey \citep[HIPASS;][]{HIPASS} and Arecibo Legacy Fast ALFA survey \citep[ALFALFA;][]{Giovanelli_2005}, have greatly advanced our understanding of \hi\ gas properties in local galaxies. These include the abundance of \hi-rich galaxies as quantified by the \hi\ mass function (HIMF; \citealt{Zwaan2005HIMF,Martin2010,JonesHIMF2018}), the dependence of galaxy clustering on \hi\ mass \citep{Meyer-2007,Martin2012,Papastergis2013,Guo2017}, the scaling relations of \hi\ mass fraction with various galaxy properties \citep{Kannappan2004,ZhangWei_2009,Li2012,HuangS2012,GASS,Obreschkow2016,xGASS,Zu2020,XiaoLi,Lu2024}, and the \hi-to-halo mass relation for both central galaxies and all the galaxies in dark matter halos \citep{Hess2013,Barnes2014,Guo2017,Paul2018,Navarro2018,Obuljen2019,Guo-HI-halo-relation,Chauhan2020,Calette2021,Chauhan2021,XiaoLi,Rhee2023,Saraf2024}.  Furthermore, the \hi\ scaling relations provide an economic but reliable way to estimate the \hi\ mass fraction for optically-selected galaxy samples that are much deeper and larger than the existing \hi\ surveys. 
This allows the abundance and clustering of \hi-selected galaxies to be measured more accurately and over larger dynamical ranges of galaxy properties. 
For instance, by applying an improved estimator to the galaxy sample from the Sloan Digital Sky Survey \cite[SDSS;][]{2000AJ....120.1579Y}, \cite{XiaoLi} obtained the first estimates of the conditional \hi\ mass functions (CHIMFs), that is, HIMFs for galaxies hosted by dark matter halos of given mass. 

The observations of \hi\ should in principle provide useful constraints on the subgrid physics implemented in hydrodynamic simulations. For instance, 
\cite{SIMBA_HI} found that SIMBA well reproduces the HIMF at $z=0$, but overestimates the \hi\ fraction of massive galaxies, implying too strong a jet-mode AGN feedback which plays a major role in regulating the \hi\ gas content in this simulation. For IllustrisTNG, the \hi\ mass of central galaxies is found to have no or weak correlation with star formation rate \citep[SFR;][]{Ma2022}, in contrast to the tight correlation between \hi\ mass fraction and SFR in real galaxies.  In \cite{XiaoLi}, the comparison of the \hi-to-halo mass relation between observations and simulations show that both Illustris and IllustrisTNG simulations significantly overpredict the \hi\ mass for both the central galaxy and all the galaxies as a whole at fixed halo mass. Due to the complex subgrid physics in the simulations, as mentioned above, it is not immediately clear what causes the agreement and disagreement between simulations and observations. 

In this work, we adopt the random forest algorithm \citep{RandomForest} to investigate the driving processes for the \hi\ gas content in IllustrisTNG (hereafter TNG) and EAGLE. Both TNG and EAGLE are state-of-the-art hydrodynamic simulations, but they implement very different feedback models, thus allowing a comparison of the effects of different feedback processes on the \hi\ contents of galaxies. Random forest is a widely-used machine learning algorithm that is able to efficiently identify the most important feature with respect to the target from a large parameter space \citep[e.g.][]{Bluck2022, Bluck2023, Baker2023, Ellison2024, Goubert2024, Jing2024, Liniu2024}. We consider galaxy properties that are previously found to be correlated with \hi\ mass fraction (e.g., specific SFR, color, stellar mass, stellar surface density, gas metallicity), as well as other properties that are potential tracers of important gas-related processes (e.g., black hole mass and accretion rate). We then compare the importance pattern of the properties in simulated galaxies with that identified in the same way from the observed galaxies in the xGASS survey, aiming to find out the driving mechanisms for the \hi\ mass fraction in low-redshift galaxies.

This paper is organized as follows. In \autoref{sec:data_simulations} we first perform the random forest analysis for the two simulations. In \autoref{sec:compare to observation} we then perform the same analysis for the xGASS galaxy sample and make comparisons between the simulations and the observational results. Finally, we summarize our work in \autoref{sec:summary}.


\section{Random Forest analysis of simulations} \label{sec:data_simulations}
\subsection{Data and Sample Selection}
\subsubsection{IllustrisTNG}
The IllustrisTNG project \citep[][hereafter TNG]{Springel_2018,Naiman_2018,Marinacci_2018,Nelson_2018,Pillepich_2018} is a suite of cosmological magneto-hydrodynamic simulations in a $\Lambda$CDM universe run with the moving-mesh code \textsc{arepo} \citep{Springel2010}. To ensure both resolution and volume, in this work we use the publicly available TNG100 data, which has a box length of 75$\hmpc$ and a baryonic mass resolution of $1.4\times 10^{6}\msun$.
IllustrisTNG adopts the cosmological parameters from \cite{Planck_2016}: $\Omega_m = 0.3089,\ \Omega_b = 0.0486,\ h = 0.6774$, and $\sigma_8 = 0.8159$. The galaxy formation and evolution models in IllustrisTNG simulations include gas cooling, star formation, galactic winds, metal enrichment, supernovae, black hole growth, active galactic nuclei (AGN) feedback \citep{Weinberger2017, Pillepich_2018}, and magnetic field. 
Dark matter halos are identified using FoF algorithm \citep{Davis_1985}. Then the \textsc{subfind} algorithm \citep{Springel_2001} is used to identify subhalos (galaxies) in FoF halos.

\subsubsection{EAGLE}
The EAGLE project \citep{Crain_2015, Schaye_2015} is a suite of hydrodynamic simulations in a $\Lambda$CDM universe. EAGLE was run using a modified version of the $N$-Body Tree-PM smoothed particle hydrodynamics (SPH) code \textsc{GADGET3}, originally developed by \cite{Springel_2005}. The simulations adopt the cosmological parameters from \cite{Planck_2014}: $\Omega_m = 0.307,\ \Omega_b = 0.0483,\ h = 0.6777$, and $\sigma_8 = 0.8288$. In this work we use the fiducial simulation Ref-L100N1504, which has a volume of side $L=100$ Mpc and a baryonic particle mass of $1.81\times 10^6 \msun$. The subgrid processes implemented in Ref-L100N1504 include 
radiative cooling, reionization, star formation, chemical enrichment, supernova feedback, black hole growth, 
and AGN feedback. Dark matter halos and subhalos are identified using the FoF and \textsc{subfind} algorithms, respectively.

\subsubsection{\hihh\ Partition Models}
Both TNG and EAGLE output only the total mass of hydrogen element in each gas particle, without providing fractions of  the ionized (\hii), atomic (\hi) and molecular (\hh) phases. The partition between \hi\ and \hh\ is done with a post-processing procedure, which is similar in the two simulations. The hydrogen gas is first divided into ionized and neutral phases, and models of the \hihh\ partition are then applied to the neutral phase to estimate the mass fraction of \hi\ and \hh\ gas. Both simulations implemented multiple models for the \hihh\ partition. 
As described in  \citet{Bahe2016}, EAGLE applied the \hihh\ models of \cite{BR06} (BR06) and \cite{GK11} (GK11). For TNG, as described in \citet{Diemer2018}, the \hihh\ partition is obtained by using the models of \cite{L08} (L08), \cite{GK11} (GK11), \cite{K13} (K13), \cite{GD14} (GD14), and \cite{S14} (S14). In most models the \hihh\ partition is based on gas surface density, which is not naturally available in hydrodynamic simulations. In TNG, the gas surface density is estimated in two different ways, either by multiplying the volume density with the Jeans length (the volumetric (vol) method), or by projecting the gas cells along the direction of the gas angular momentum (the projected (map) method). In 
EAGLE, the \hi\ and \hh\ gases are assumed to have the same scale height so that the ratio of surface densities of \hh\ and \hi\ is the same as the ratio of their volume densities.

\begin{figure*}[ht]
    \centering
    \includegraphics[width=\textwidth]{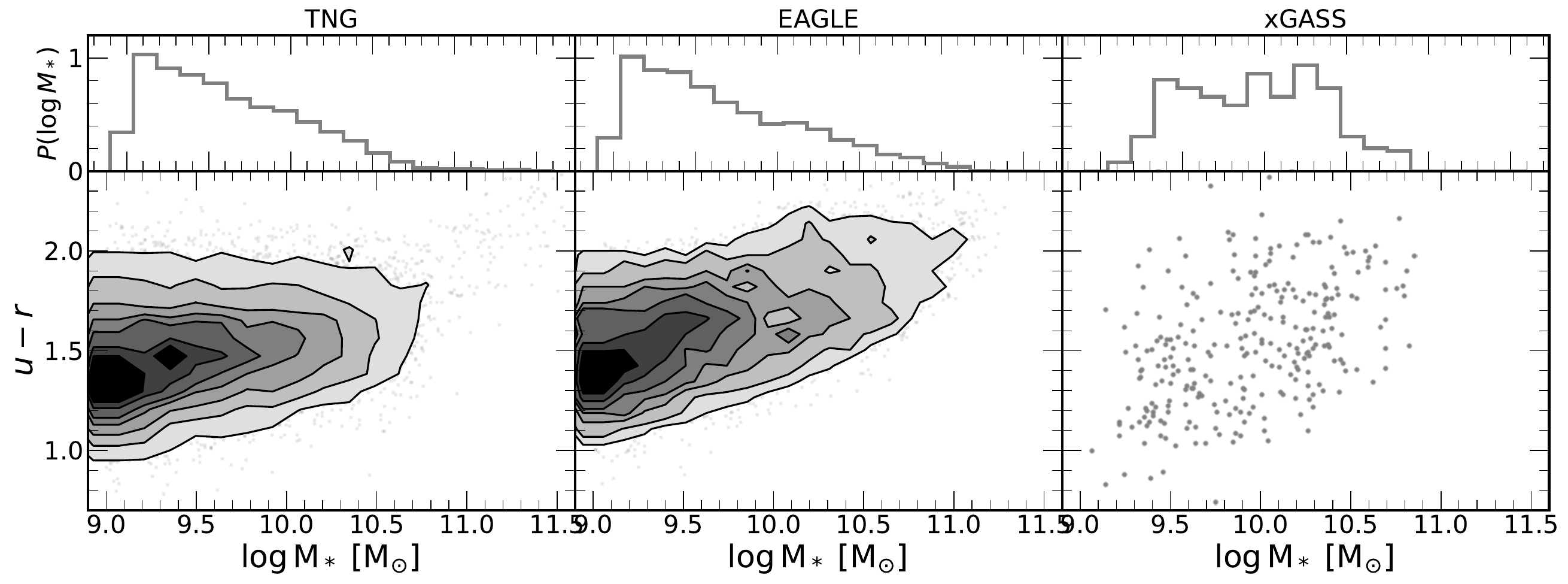}
     \caption{\textbf{Bottom panel: }The color-mass diagram of TNG, EAGLE, and xGASS sample used in this work. The grey contours show the distribution of simulation samples. Contours from inner-most to outer-most include 10\%, 25\%, 40\%, 55\%, 70\%, 85\%, and 95\% of the total sample. Galaxies outside the outer-most contour are denoted as small grey dots. \textbf{Upper panel: }The normalized stellar mass distribution of TNG, EAGLE, and xGASS sample.}
     \label{fig:color_mass}
\end{figure*}

\subsubsection{Galaxy Properties}\label{sec:sim_properties}

We consider the following galaxy properties for the random forest analysis of the simulations:

\begin{itemize}
    \item $\logmstar$: 
    the logarithm of the stellar mass, in units of solar mass. This is the sum of masses of all stellar particles within 30 kpc. A fixed aperture of 30 kpc is demonstrated to track the Petrosian radii used in observations \citep{Schaye_2015}.
    \item $\log \mu_*$: 
    the logarithm of the stellar surface mass density, defined by $\mu_\ast\equiv\mstar/(2\pi R_{e}^2)$, where $R_{e}$ is the half stellar mass radius in units of kpc.
    \item \logssfr: 
    the logarithm of the specific star formation rate, defined as $\rm \log SFR/M_*$, where SFR is the instantaneous star formation rate calculated using all the star-forming gas elements belonging to the subhalo for EAGLE galaxies and using star-forming gas cells within 2 times of the half stellar mass radius for TNG galaxies.
    \item $\log \sigma_*$: 
    the logarithm of the stellar velocity dispersion. For EAGLE galaxies, $\sigma_\ast$ is the one-dimensional velocity dispersion of all stellar particles. For TNG galaxies, $\sigma_\ast$ is approximated using the one-dimensional velocity dispersion of all subhalo member particles.
    \item $u-r$: color index defined by the difference in absolute magnitude between $u$ and $r$ band.  The absolute magnitudes are calculated using stellar population synthesis models with dust obscuration taken into account \citep{Trayford2015, Nelson_2018}.
    \item $\rm \log(O/H)+12$: gas phase metallicity of star forming gas.
    This is calculated using all the star-forming gas elements belonging to the subhalo for EAGLE galaxies and star-forming gas cells within 2 times of the half stellar mass radius for TNG galaxies.
    \item \logmbh: the logarithm of the mass of the central supermassive black hole, in units of $\msun$.
    \item \logmbhdot: the logarithm of the accretion rate of the central  supermassive black hole, in units of $\rm \msun/yr$.
    \item $\rm \log V_{max}$: the logarithm of the maximum value of the spherically-averaged rotation velocity, in units of $\rm km/s$.
    \item $\log Z_*$: the logarithm of the stellar metallicity, derived from all stellar member particles in EAGLE, and from stellar member particles within twice the half stellar mass radius in TNG.
    
\end{itemize}

We note that, in both EAGLE and TNG, the \hi\ mass is measured using an aperture of 70 kpc, and the stellar mass is measured within 30 kpc for the consistency of sample selection. For all other properties, we directly take the measurements from the subhalo catalogs, publicly available through the websites of the two simulations\footnote{EAGLE: https://icc.dur.ac.uk/Eagle/database.php; TNG: https://www.tng-project.org/data/}. For the star formation rate and gas metallicity, EAGLE considers all gas particles belonging to the subhalo, while TNG only considers gas cells belonging to the subhalo within twice the half stellar mass radius ($\rm R_e$). Star-forming gas is expected to concentrate in the inner region, mostly within $\rm 2R_e$. In addition, the feature importance analysis is sensitive to the ranking but not the absolute value of the properties considered. For these reasons, the small differences in property definitions are expected to have negligible effects on the results presented in this work.

\subsubsection{Sample Selection}

Our investigation focuses on star-forming central galaxies, for the following two considerations. First, most of the quenched galaxies do not have reliable \hi\ gas measurements due to the limited mass resolution of the simulations. Second, the \hi\ content of satellite galaxies is known to suffer from environmental effects such as ram-pressure stripping and tidal stripping, making it hard to reliably assess the importance of properties internal to the galaxies. We will come back and study the effect of environments in the future. Here, we select star-forming central galaxies with stellar mass $\mstar > 10^9 \msun$ and specific star formation rate $\logssfr > -11$. To avoid unreliable \hi\ gas meaurements, we further require the \hi-to-stellar mass ratio to be larger than a fixed limit: $\lgfhi > -2$. These restrictions give rise to a sample of 10,205 galaxies from TNG, and a sample of 6,287 galaxies from EAGLE. The left two panels of \autoref{fig:color_mass} show 
the distribution of sample galaxies in the $u-r$ versus \logmstar\ diagram. As expected, galaxies in both samples are 
predominantly blue, with $u-r<2.2$. Both samples are dominated by galaxies of low to intermediate masses, with very few galaxies 
with mass exceeding $10^{11}\msun$. EAGLE galaxies show a positive correlation between the color and the mass, 
which is not obviously seen in TNG. 

\begin{figure}[t!]
    \centering
    \includegraphics[width=0.48\textwidth,height=0.4\textwidth]{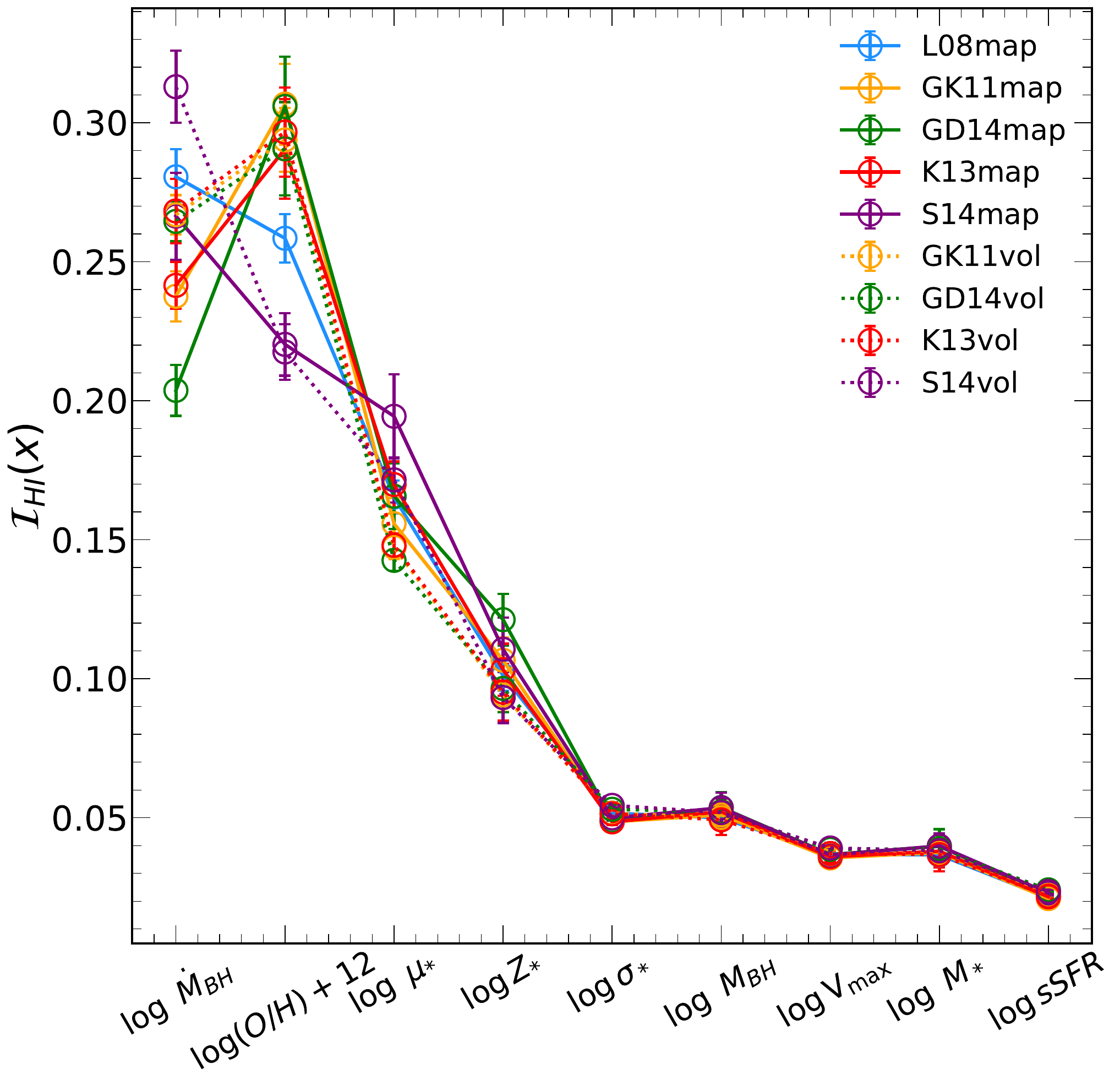}
    \caption{Feature importance of the TNG sample. The solid/dashed lines represent projected(map)/volumetric(vol) method to estimate the gas surface density. \hihh\ partition models are denoted by different colors (blue: \cite{L08}, orange: \cite{GK11}, green: \cite{GD14}, red: \cite{K13}, purple: \cite{S14}).}
    \label{fig:I_allparams_HIH2}
\end{figure}


\subsection{Feature Importance Ranking in the Simulations} \label{sec:HIH2_partition_model_compare}

Following common practice, we use the Gini importance value provided by the random forest regressor to assess the significance of individual galaxy properties  to the \hi-to-stellar mass ratio. The properties used are listed and described in \autoref{sec:sim_properties}. We use the \textsc{scikit-learn} \footnote{https://scikit-learn.org/stable/index.html} \citep{scikit-learn} machine learning python package to realize the random forest algorithm. The importance of properties for the TNG sample is shown in \autoref{fig:I_allparams_HIH2} in a decreasing 
order. Plotted in different colors are results for different \hihh\ partition models, while the solid and dashed lines of the same color represent results for a given \hihh\ partition model using two different methods to determine the gas surface 
density (see \autoref{sec:sim_properties}). 
The uncertainties of the feature importance are estimated by the scatter between 40 subsamples, each constructed by 
randomly selecting two thirds of the galaxies from the full sample.

\begin{figure}[t!]
    \centering
    \includegraphics[width=0.5\textwidth,height=0.4\textwidth]{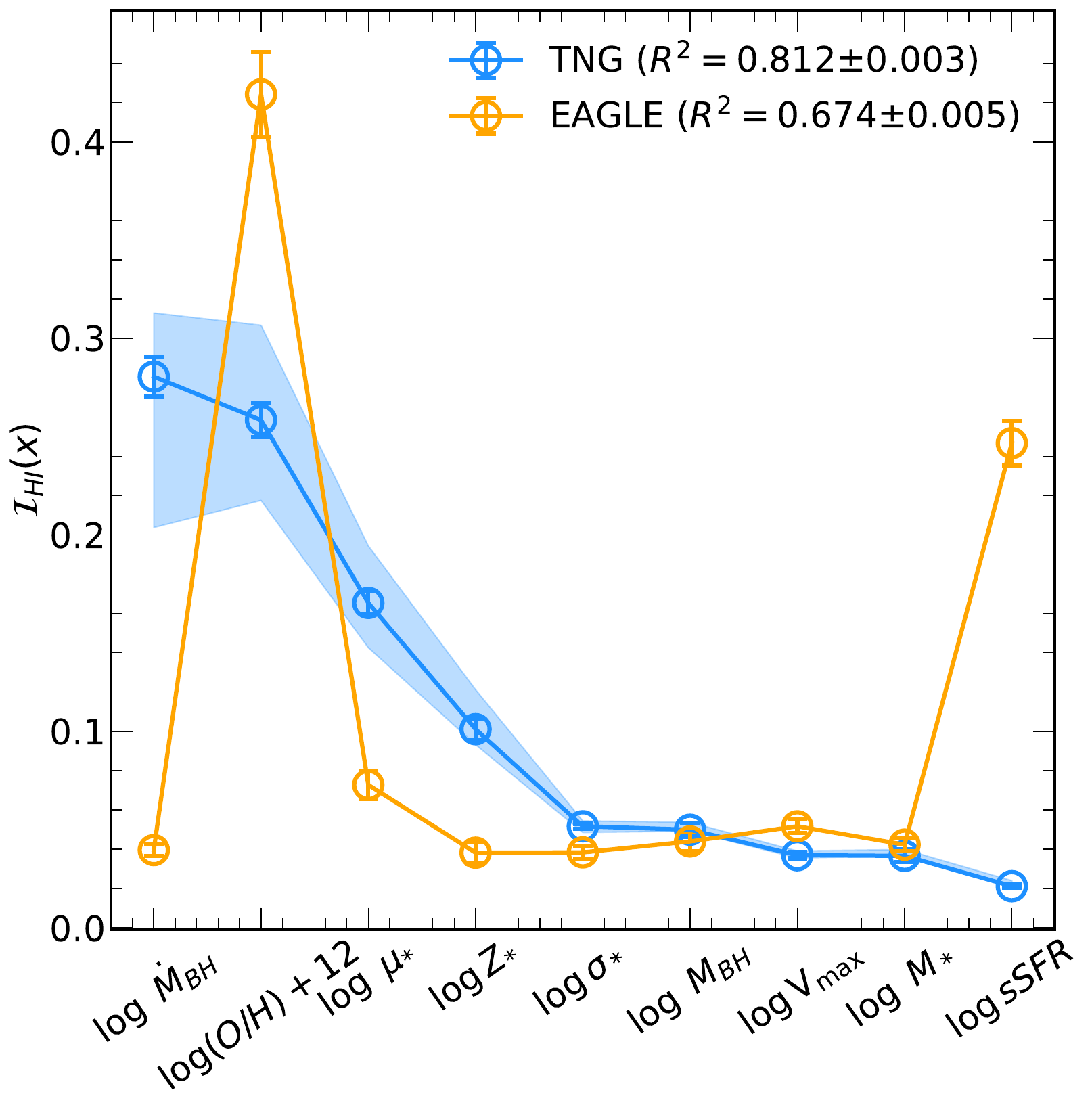}
    \caption{Feature importance of the EAGLE(orange) and TNG(blue) sample. The shaded region represents the uncertainty of feature importance due to \hihh\ partition models. The $R^2$ score of each sample is shown in the legend. }
    \label{fig:I_TNG_EAGLE}
\end{figure}

\begin{figure*}[ht]
     \centering
     \includegraphics[width=0.99\textwidth]{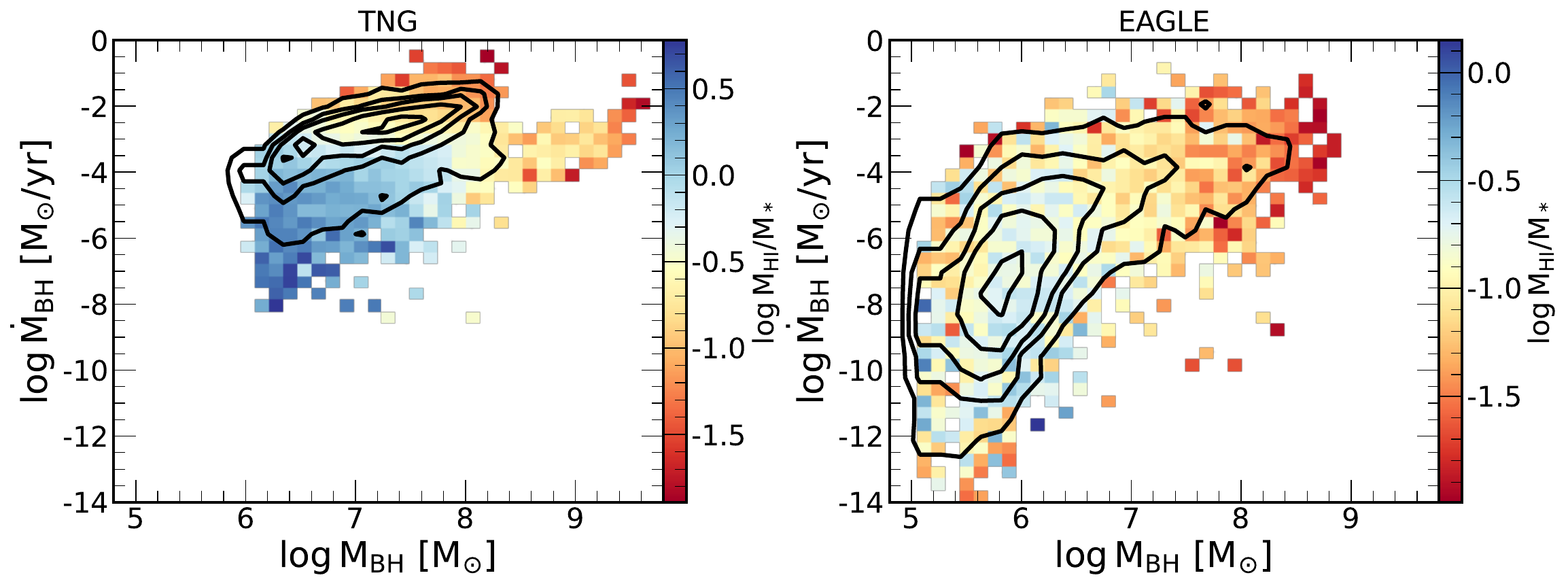}
     \caption{Black hole accretion rate as a function of black hole mass for the TNG(left) and EAGLE(right) sample, color-coded by \hi\ fraction. The black contours show the number density distribution of the sample.}
     \label{fig:MBHdot_MBH_fHI}
\end{figure*}

In all the models, the black hole accretion rate (\logmbhdot) and the gas-phase metallicity (\logoh) are ranked as the top two properties in terms of feature importance, with \logmbhdot\ the most important for L08map, S14vol and S14map, and \logoh\ the most important 
for all other models. The two properties are followed by the stellar surface mass density ($\mu_\ast$) and the 
stellar metallicity ($Z_\ast$). The specific star formation rate, \logssfr, is surprisingly ranked as the least 
important property, a result that is opposite to the well-known correlation between cold gas content and star formation in galaxies. Nevertheless, the results shown in \autoref{fig:I_allparams_HIH2} demonstrate that the ranking of feature importance obtained 
from TNG is robust to the choice of \hihh\ partition models for all the properties except between \logmbhdot\ and \logoh. 

For EAGLE, although two \hihh\ partition models (BR06 and GK11) were used, the resulting \hi\ data is publicly available 
only for BR06. In \autoref{fig:I_TNG_EAGLE}, the feature importance obtained from the BR06 model are plotted as the orange circles 
with error bars. For comparison, the results of the different \hihh\ partition models from TNG are plotted as the blue shaded region, 
with the upper and lower envelopes corresponding, respectively, to the maximum and minimum importance among 
all the \hihh\ partition models 
for any given property. The blue circles with error bars represent the L08map model in TNG, which estimates the 
ratio between \hh\ and \hi\ gas surface density as a power law of midplane pressure, in the same way as the BR06 model. 
In what follows we will only consider the L08map model for TNG and the BR06 model for EAGLE, for simplicity.



As can be seen from \autoref{fig:I_TNG_EAGLE}, the most important property in EAGLE is \logoh, followed by \logssfr\ and $\mu_\ast$. 
This is in contrast to TNG, where \logmbhdot\ is ranked the top and \logssfr\ the bottom. The importance of \logmbhdot\ in EAGLE is nearly zero. 
The only similarity between the two simulations is that \logoh\ is important in both. The overall difference between the two 
simulations is quite striking: while the HI content in EAGLE is dominated by two properties, 
in TNG more factors seem to be involved. This indicates that EAGLE and TNG are intrinsically different. 
In the following subsection, we examine sources leading to the difference between the two simulations 
as revealed by the feature importance of HI contents of galaxies, and show that the difference 
can be understood in terms of the different treatments of key subgrid processes that regulate 
star formation and feedback.


\subsection{Implications of the Feature Importance} \label{subsec:discussion_interpretation}

We first examine the correlation of the \hi\ gas content with the black hole accretion rate. 
This rate is one of the most important properties in TNG but shows no importance in EAGLE. 
In \autoref{fig:MBHdot_MBH_fHI}, we show the distribution of galaxies   
in the $\mbhdot$ versus black hole mass ($\mbh$) plane, color-coded by the \hi-to-stellar mass ratio 
(hereafter \hi\ mass fraction or $\hi/\mstar$). As one can see, EAGLE shows an overabundance of low mass black 
holes ($\mbh < 10^6 \msun$) in comparison to TNG, possibly caused by the different black hole seed mass 
adopted in TNG ($8\times 10^{5}\hmsun$) and EAGLE ($10^{5}\hmsun$). Both simulations show a positive 
correlation between \logmbhdot\ and \logmbh. In TNG and below $\mbh \sim 10^{8.2} \msun$, the \hi\ mass 
fraction is negatively correlated with \logmbhdot\ and nearly independent of black hole mass. 
Above $\mbh \sim 10^{8.2} \msun$, the \hi\ mass fraction plummets due to the onset of the kinetic mode 
of AGN feedback. Such a feedback prevents the cooling of gas in the CGM and the replenishment of the 
cold gas in ISM, leading to the quenching of star formation \citep{Zinger2020, Piotrowska2022}. 
In contrast, EAGLE shows only a weak negative correlation between \logmbh\ and \hi\ mass fraction, and 
the value of \logmbhdot\ is on average much lower than that in TNG. These results are consistent 
with the feature importance as found above.


  \begin{figure*}[ht]
     \centering
     \includegraphics[width=0.99\textwidth]{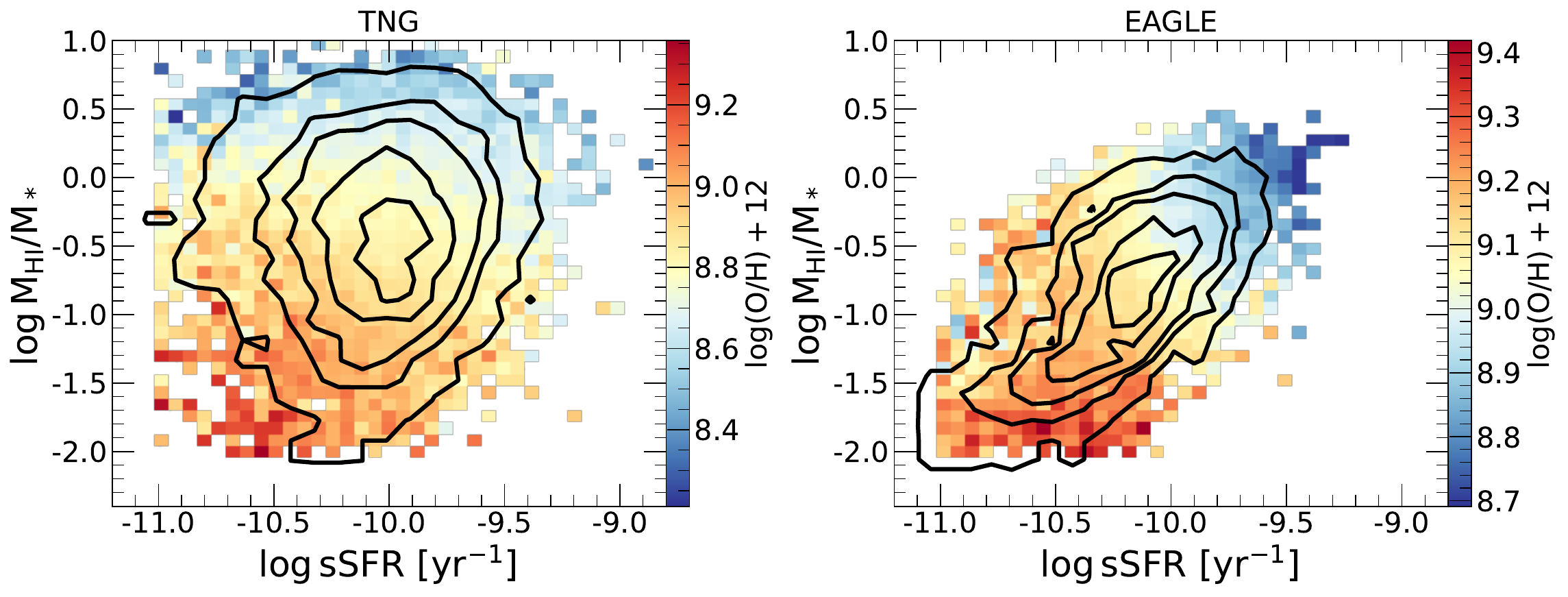}
     \caption{\hi\ fraction as a function of specific star formation rate for the TNG(left) and EAGLE(right) sample, color-coded by gas phase metallicity of star-forming gas. The black contours show the number density distribution of the sample.}
     \label{fig:fHI_logssfr_logoh}
 \end{figure*}
 
In EAGLE, most of the sample galaxies have $\mstar < 10^{10.5}\msun$, corresponding to a halo mass of $\mhalo \lesssim 10^{12}\msun$ based on the stellar-to-halo mass relation of central galaxies at z=0 \citep{2018ARA&A..56..435W}. In these low-mass halos, the gas outflow driven by stellar feedback is able to escape from the host halo and to affect the CGM \citep{Mitchell2020a,Wright2024}. The rate of gas inflow into the ISM is thus very low, 
leading to the low rate of black-hole accretion observed in \autoref{fig:MBHdot_MBH_fHI} and  
inefficient AGN feedback. In addition, since the AGN feedback energy in EAGLE is not released until it is high 
enough to increase the temperature of a neighboring gas particle by $\rm \Delta T_{AGN}=10^{8.5}K$ 
\citep{Schaye_2015,Crain_2015}, the feedback energy released at one time does not necessarily 
correspond to the accretion rate of the supermassive black hole at the same time, which 
may reduce the correlation between the cold gas content and the accretion rate of the central black hole.    
Even for galaxies with high black hole accretion rates, EAGLE does not show a significant correlation 
between the black hole accretion rate and the \hi\ mass fraction, 
indicating that it is the pulse-like characteristic, not the low accretion rate,  
that is responsible for the low importance of the black hole accretion rate in EAGLE. 
This is consistent with the result of \citet{Ward2022} who found 
a weak trend of the molecular gas fraction with the luminosity of the central AGN in EAGLE and interpreted it as 
a consequence of the ``pulsed" AGN feedback. 

In contrast, most galaxies in the TNG sample are in the thermal mode of AGN feedback which continuously inject 
energy into the surrounding ISM, producing a strong negative relationship between the \hi\ mass fraction and the 
black hole accretion rate. In addition, as shown in \cite{SIMBA_HI}, the amplitude of the HIMF decreases in TNG but 
increases in SIMBA and EAGLE as one goes from $z=0$ to $z=2$. 
Since both TNG and SIMBA implement stellar feedback using decoupled kinetic winds, 
the striking difference in the evolution of the HIMF is likely caused by the thermal AGN feedback 
implemented in TNG, which is expected to be stronger at $z=2$ than at $z=0$ \citep{Weinberger2017}. 
\cite{Ma2022} and \cite{Ward2022} also found a notable negative correlation between the cold gas fraction 
and the accretion rate of the thermal mode AGN.
These previous studies and our results all show that the thermal AGN feedback in TNG
affects significantly the cold gas content in star-forming galaxies at $z=0$. 


Next, we examine the correlation of \hi\ gas content with the gas-phase metallicity and sSFR. As seen from \autoref{fig:I_TNG_EAGLE}, \logoh\ is identified as a very important feature in both EAGLE and TNG, while \logssfr\ is only important in EAGLE. \autoref{fig:fHI_logssfr_logoh} plots the \hi\ mass fraction as a function of \logssfr, with the black contours showing the number density of galaxies and the color-coding showing \logoh. 
As can be seen from the right panel, EAGLE galaxies show strong correlations between each pair of \hi\ mass fraction, sSFR and gas-phase metallicity, a result that is in broad agreement with previous results\citep{Ellison2008,Mannucci2010,Bothwell2013,Hughes2013,Zahid2014,DeRossi2017}. As pointed out in \cite{Lagos2016}, these correlations are a consequence of the self-regulation of star formation, which dominates the gas-star-gas cycling for low-mass galaxies like the EAGLE galaxies studied here. In this case, the inflow of pristine gas reduces gas metallicity and triggers star formation, which in turn release feedback energy and enriched metals to the surrounding gas, generating gas outflows and suppressing gas inflows. Galaxies thus evolve in a quasi-static way so that gas inflow, gas outflow and star formation occur at balanced rates, producing a tight relation 
between star formation and cold gas mas. 
For massive galaxies the self-regulation of star formation is broken due to AGN feedback, driving significant deviation from the relations between gas mass, stellar mass and gas metallicity \citep{Zerbo2024}. However, many of these 
massive galaxies are quenched galaxies that are not selected into 
our sample of star-forming galaxies. 

\begin{figure*}[ht]
     \centering

     \includegraphics[width=0.485\textwidth]{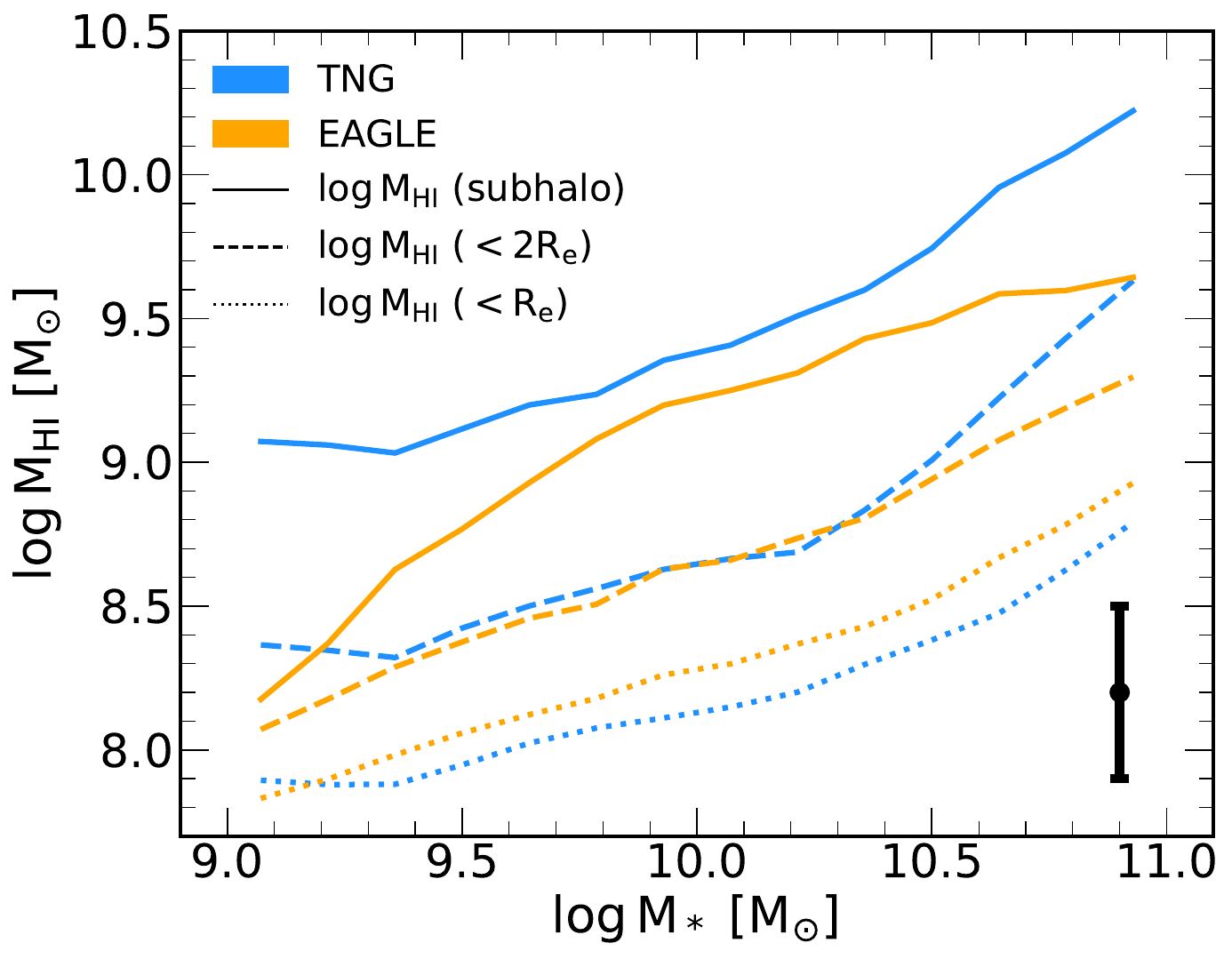}
     \includegraphics[width=0.50\textwidth]{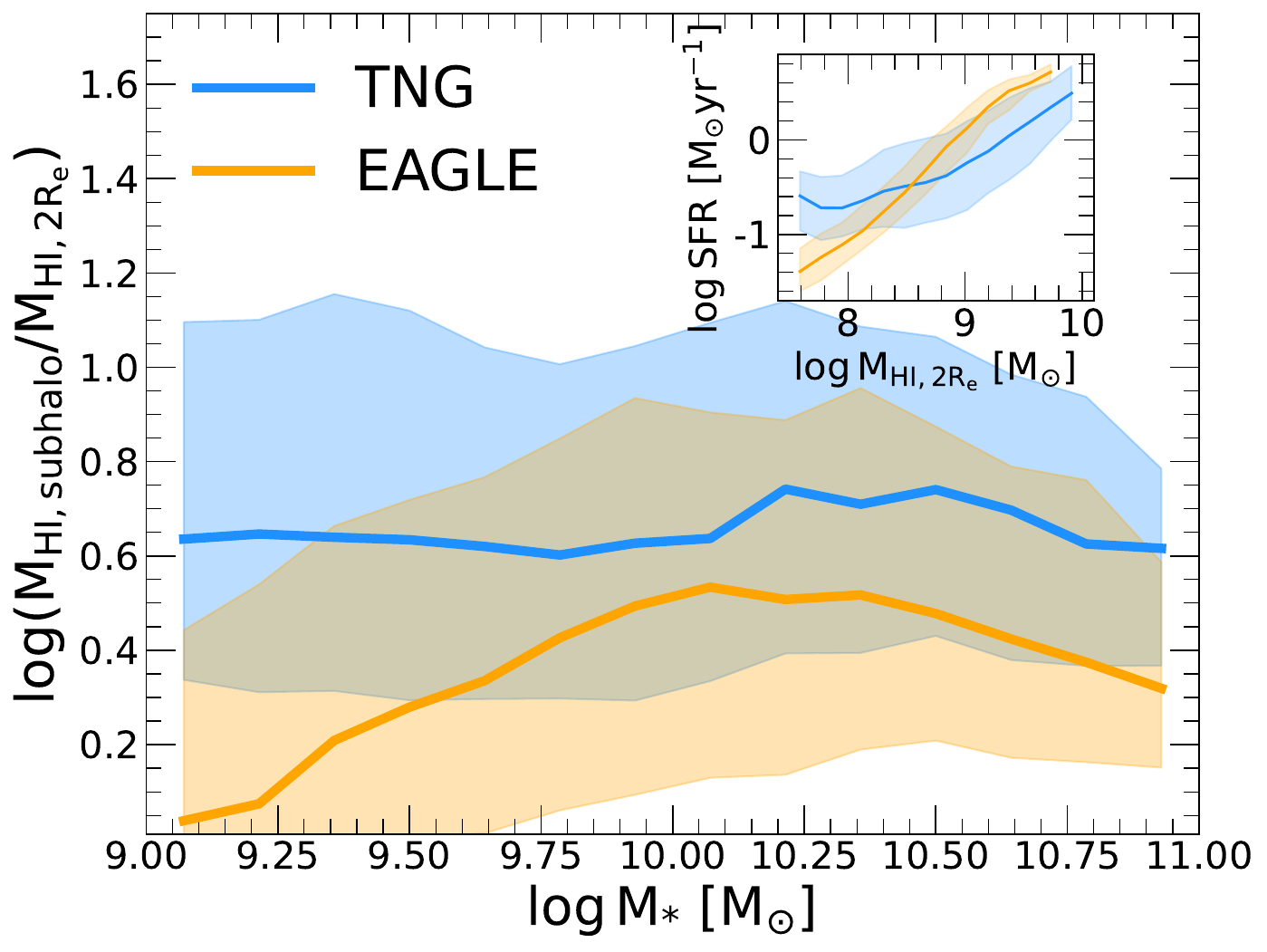}
     \caption{\textbf{Left: }The \hi\ mass enclosed within $R_e$(dotted), $2R_e$(dashed), and the radius of the host subhalo(solid) as a function of stellar mass in TNG(blue) and EAGLE(orange). The black error bar represents the typical error for clarity. \textbf{Right: }The ratio of the total \hi\ mass within the host subhalo to the \hi\ mass within $2R_e$ as a function of stellar mass. The shaded region represents the $1\sigma$ scatter. The inset panel shows the star formation rate as a function of the \hi\ mass enclosed within $2R_e$.}
     \label{fig:gas_mass_ratio}
 \end{figure*}

Unlike EAGLE and contrary to observations, the \hi\ mass fraction of TNG galaxies is correlated only with gas metallicity, and shows no or weak correlation with sSFR.
We find that the different behaviors of the two simulations are caused by the different fractions of \hi\ gas that is distributed outside the galaxies. This can be seen from \autoref{fig:gas_mass_ratio}. The left-hand panel shows the \hi\ mass enclosed within $R_e$, $2R_e$ and the radius of the host subhalo as a function of stellar mass, for both simulations. Within $R_e$ and $2R_e$, the two simulations show similar \hi-to-stellar mass relations. When the total \hi\ gas is included, TNG galaxies have significantly larger amounts of gas outside the stellar radii, indicating more extended \hi\ discs relative to the stellar disk. The right-hand panel further shows the ratio of the total \hi\ mass within the host subhalo to the \hi\ mass within 2$R_e$, which quantifies how centrally concentrated the \hi\ gas is. The ratio is roughly constant at $\sim 4.5$ in TNG, indicating that only one fifth of the \hi\ gas is locked within galaxies. In contrast, the \hi\ gas in EAGLE is more confined within galaxies, with much smaller values of the ratio at all masses. In particular, at lowest masses ($\mstar \sim 10^9\msun$) nearly all the \hi\ gas in EAGLE galaxies is located within $2R_e$. 
Considering the SFR is calculated within $2R_e$, this result indicates that only a small fraction of \hi\ gas contributes to star formation in TNG, thus resulting in a weak correlation between \hi\ gas fraction and sSFR. If one only considers the \hi\ gas within $\rm 2R_e$, as shown in the inset, the \hi\ mass is indeed positively correlated with SFR in both simulations, although the relation in TNG shows a flatter slope and a larger scatter compared to that in EAGLE.

In summary, the above results combine to show that the \hi\ contents in EAGLE and TNG are regulated by distinct feedback processes. In EAGLE and for galaxies of $\mstar < 10^{10.5}\msun$, stellar feedback efficiently expels ISM gas out of host halos, leading to suppressed rates of gas inflow onto both ISM and central supermassive black hole \citep{Mitchell2020a,Davies2020} and thus a negligible effect of AGN feedback. In this case, star formation is self-regulated, leading to balance between gas outflow/inflow and star formation, and a tight correlations between SFR and \hi\ mass fraction and stellar mass. AGN feedback is important in EAGLE only for massive halos ($\mhalo > 10^{12}\msun$) where the gravitational potential is deep enough to prevent the star formation-driven outflow \citep{Bower2017}, and so the established relation between the HI mass and the star formation rate is not destroyed 
by the AGN feedback in low-mass galaxies that dominate our sample. 

In TNG, galactic wind driven by star formation is gentle and mostly stays within the host halo. The gas returns to the ISM and CGM after cooling \citep{Davies2020,Ayromlou2023}, forming a large and spatially extended \hi\ reservior \citep{Grand2019,Diemer2019,Yang2024arxiv}. The \hi-to-stellar mass ratio is weakly correlated with sSFR which is determined by the gas within $2R_e$ of the galaxy. The central supermassive black hole is able to maintain a high accretion rate, continuously heating surrounding gas via thermal AGN feedback and resulting in a strong correlation between the \hi\ fraction and the black hole accretion rate discussed above.  
In more massive halos ($\mhalo > 10^{12}\msun$), AGN feedback is dominated by the kinetic mode, which can significantly reduce the CGM gas reservoir. 
In this case, the amounts of star formation and cold gas are both suppressed, 
leading to the formation of quenched galaxies that are not included 
in the samples of star-forming galaxies concerned here.    

Despite of the very different feedback models implemented 
in the two simulations and their different predictions for the 
relation between cold gas mass and star formation rate, both 
simulations predict a strong correlation between the gas-phase metallicity 
and the HI mass fraction. Such a tight relation is expected if 
metals produced by star formation is well mixed in the ISM. In this case,
the fractional metal loss is proportional to the fractional 
total-mass loss from the ISM and the ISM metallicity is determined by 
the remaining gas fraction even when outflows and the inflow of 
low-metallicity gas are involved \citep[e.g.][]{LuMoLu2015}.    
Thus, the tight correlation between the gas metallicity and gas fraction 
seen in both simulations is likely a result of the subgrid physics 
that leads to a roughly uniform mixing of metals with the ISM before 
feedback effects drive gas out from galaxies. The strong (weak) correlation
between the gas metallicity and the specific star formation rate 
can then be understood as a result of the tight (loose) relation 
between the cold gas fraction and the specific star formation rate
in EAGLE (TNG) discussed above.

\section{Comparison with Observations} \label{sec:compare to observation}


\subsection{Observational Sample and Galaxy Properties}

\subsubsection{The xGASS sample}

The GALEX Arecibo SDSS survey \citep[GASS;][]{GASS}
is a targeted \hi\ survey observed with the Arecibo telescope for a sample of galaxies with redshift $0.025<z<0.05$
and a flat stellar mass distribution in the range $10^{10}\msun<\mstar<10^{11.5}\msun$.
The GASS sample is randomly selected from a parent sample of ~$12,000$ galaxies located in the overlapping region
among SDSS data release 6 \citep[][]{Adelman_McCarthy_2008}, GALEX \citep[][]{Martin_2005} Medium Imaging Survey,
and the ALFALFA survey footprint. Each galaxy is observed with Arecibo until its \hi\ 21cm emission line is detected
or an upper limit of $1.5\%$ of the \hi-to-stellar mass ratio ($\fhi$) is reached.
The xGASS entends the GASS survey down to a stellar mass lower limit of $10^9 \msun$, by further
observing a sample of galaxies with $10^{9}<\mhi<10^{10.2}$ and $0.01<z<0.02$.
Here we use the \textit{xGASS representative sample} constructed by \cite{xGASS}, which includes 1179 galaxies and is representative to the general galaxy population. From this sample, we select central galaxies using the SDSS DR7 group catalog \citep{Yang2007-group-catalog}, and exclude galaxies with $\logssfr < -11$ or without gas metallicity measurements (details of $\logssfr$ and gas metallicity are described in \autoref{subsec:obs_properties}). The remaining sample contains 278 galaxies with \hi\ detections. We will use this sample for our 
Random Forest analysis. As shown in the right panel of \autoref{fig:color_mass}, the xGASS sample has a flat stellar mass distribution and locates in the star-forming sequence with a slope similar to that of the EAGLE sample. 

 \subsubsection{Galaxy properties} \label{subsec:obs_properties}

We consider the following five properties for the random forest analysis of the xGASS sample.
\begin{itemize}
    \item $\logmstar$: logarithm of stellar mass, taken from the NASA-Sloan Atlas \footnote{http://nsatlas.org} \citep[NSA;][]{2011AJ....142...31B}, estimated by performing a spectral energy distribution (SED) fitting to the SDSS photometry, assuming a Chabrier stellar initial mass function \citep{Chabrier2003}. 
    \item $\log \mu_*$: logarithm of stellar surface density, defined as $\log \mu_* \equiv \log \mstar/(2\pi R_{50}^2)$, where $\rm R_{50}$ is the elliptical Petrosian half light radius taken from the NSA.
    \item $u-r$: color index defined by the $u$ band and $r$ band absolute magnitude. The absolute magnitudes are taken from the NSA, measured within an elliptical Petrosian aperture based on GALEX and SDSS images respectively, with Galactic extinction corrected.
    \item $\rm \log (O/H)+12$: gas phase metallicity, taken from the MPA/JHU SDSS catalog \footnote{https://wwwmpa.mpa-garching.mpg.de}, derived from SDSS optical spectra using all the most prominent emission lines \citep{Tremonti2004}. 
    \item $\log \sigma_*$: logarithm of central stellar velocity dispersion, taken from the MPA/JHU SDSS catalog.
\end{itemize}

\begin{figure}[t!]
     \centering
     \includegraphics[width=0.45\textwidth,height=0.35\textwidth]{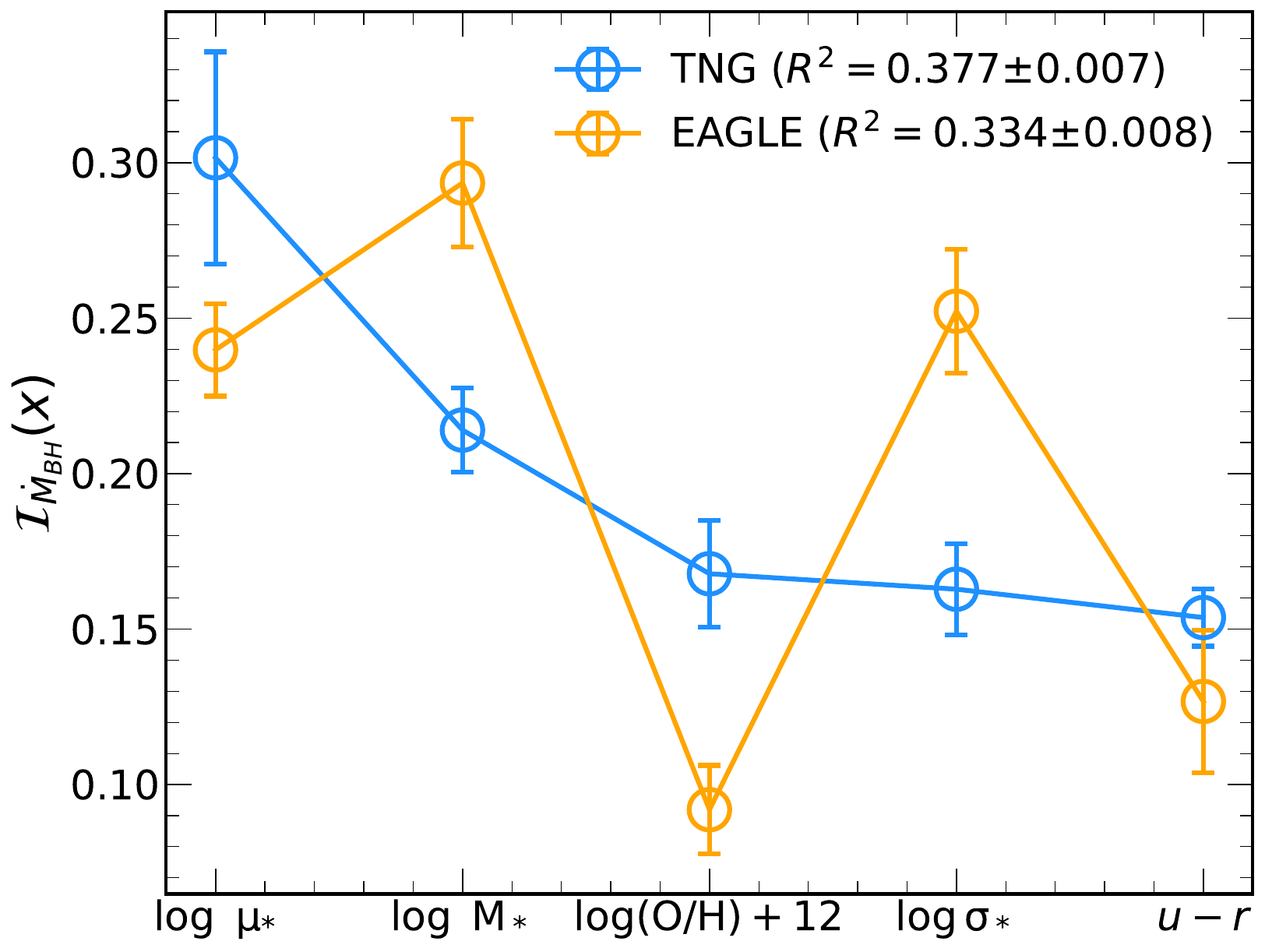}
     \caption{Feature importance with respect to black hole accretion rate in TNG sample.}
     \label{fig:Mbhdot_correlation}
 \end{figure}

As mentioned above, $\rm\mu_\ast$ and $u-r$ are found to be tightly correlated with \hi\ mass fraction in previous studies of \hi\ scaling relations \citep[e.g.][]{Kannappan2004,ZhangWei_2009,XiaoLi}. The other properties, \mstar, \logoh\ and $\sigma_\ast$, are included in order for comparison with the simulations. Due to the lack of measurements, the rest of the properties considered above for the analysis of the simulations are not directly included here. However, some of them are indirectly included through their correlations with the 
properties selected. These include the sSFR, the stellar metallicity and the black hole mass, which are known to be tightly correlated with $u-r$, $\rm \log (O/H)+12$, and $\rm\sigma_\ast$, respectively. This is also the case for $\log V_{\rm max}$, which is tightly correlated with \logmstar\ according to the well-established stellar-to-halo mass relation of central galaxies in galaxy groups/clusters \citep[][and references therein]{2018ARA&A..56..435W}. For the black hole accretion rate ($\dot{M}_{\rm BH}$), it is not immediately clear whether its importance is included indirectly.   
In order to figure this out, we examined the feature importance of the five properties to $\dot{M}_{\rm BH}$ with the random forest technique. As can be seen from \autoref{fig:Mbhdot_correlation}, for TNG the surface mass density ($\mu_\ast$) presents the highest importance to  $\dot{M}_{\rm BH}$, while \mstar, $\mu_\ast$ and $\sigma_\ast$ show similarly high importance in EAGLE. This result implies that, when $\dot{M}_{\rm BH}$ is not included in the random forest analysis, its importance may be inferred from the importance of $\mu_\ast$ for TNG and the combination of \mstar, $\mu_\ast$ and $\sigma_\ast$ for EAGLE. One should keep this result in mind when 
interpreting the results presented in the following. 

\subsection{Feature Importance to the \hi\ Fraction}

\begin{figure}[t!]
    \centering
    \includegraphics[width=0.45\textwidth,height=0.35\textwidth]{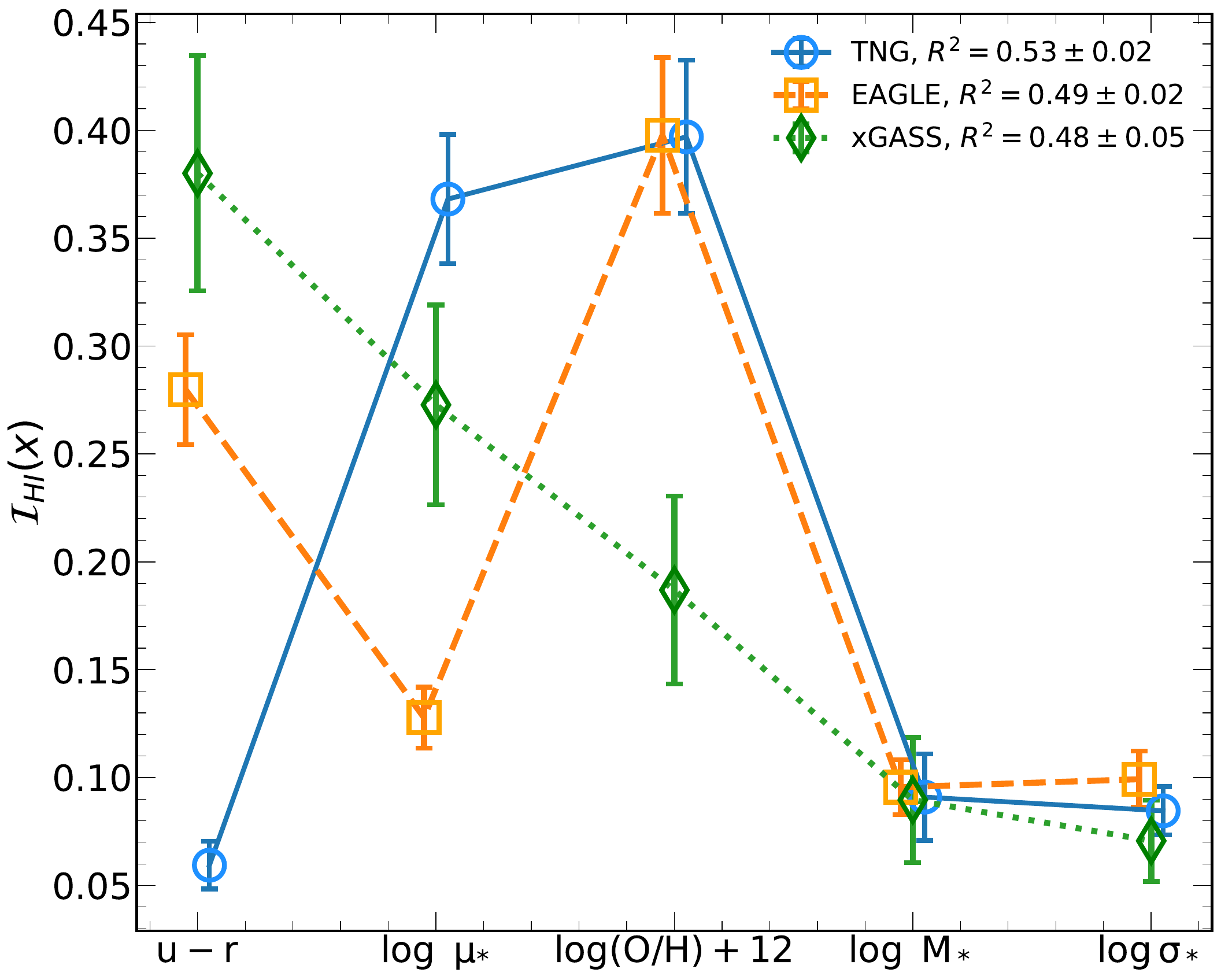}
     \caption{Feature importance of xGASS(green dotted), TNG(blue solid), and EAGLE(orange dashed). For simulation data, the error bar is derived from 40 subsamples that have the same stellar mass distribution as xGASS sample, and the measurement error of $u-r$, $\log \mu_*$, and $\rm \log (O/H)+12$ are considered. A small horizontal shift is added to the lines to avoid overlap.}
     \label{fig:I_xgass}
\end{figure}

We perform random forest analyses to determine the importance of the five galaxy properties described above to the \hi\ mass fraction, for both the observational sample xGASS and the EAGLE and TNG simulations. The results are shown in \autoref{fig:I_xgass}. For real galaxies in xGASS, as can be seen, the two most important features are $u-r$ and $\mu_*$. Both \mstar\ and $\sigma_\ast$ are ranked quite low, while \logoh\ is in the middle. To make the comparison between simulations and the observation more meaningful, we have trimmed the EAGLE and TNG samples so that they have the same distributions in stellar mass as xGASS. For each simulation we repeat the procedure of sample trimming and the random forest analysis for 40 times. The final importance of each feature and its error are given by the average and the scatter of the 40 subsamples. Furthermore, we have added a ``measurement error'' to $u-r$, $\log \mu_\ast$ and \logoh\ in the simulation data, assuming a Gaussian distribution with a full width at half maximum of 0.15, 0.05 and 0.15, respectively. These values are the typical measurement errors in the xGASS sample. We do not attempt to add errors for \logmstar\ and $\rm \log \sigma_\ast$ as their importance are too low to affect the result significantly. It should be pointed out, though, that the ``errors'' added here only represent the lower limits, 
as real data have more uncertainties (e.g. in the metallicity and in the stellar initial mass function) that are difficult 
to model reliably. Fortunately, our results remain unchanged even if we completely ignore the errors, suggesting that 
the results are robust to the uncertainties in the property measurements. 

\begin{figure*}
     \centering
     \includegraphics[width=\textwidth]{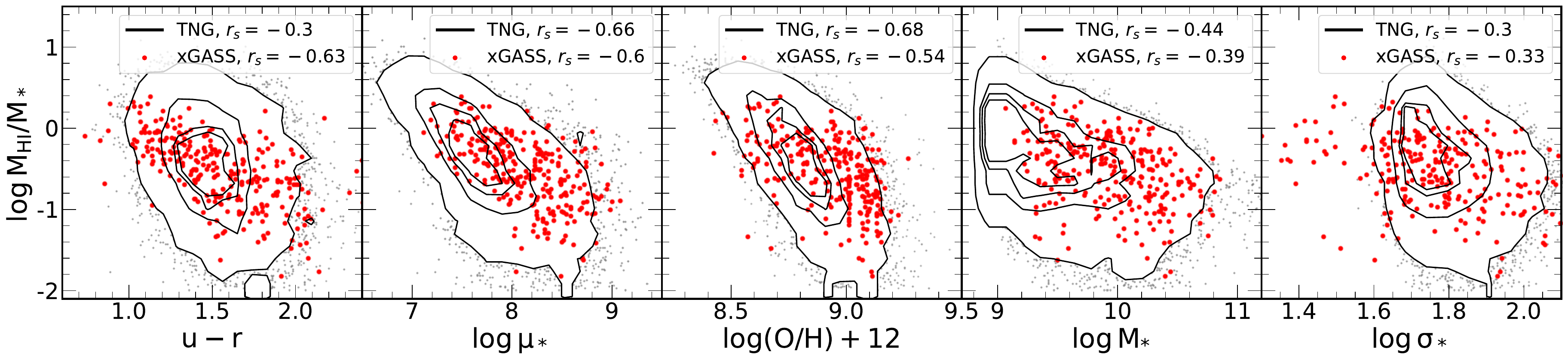}
     \includegraphics[width=\textwidth]{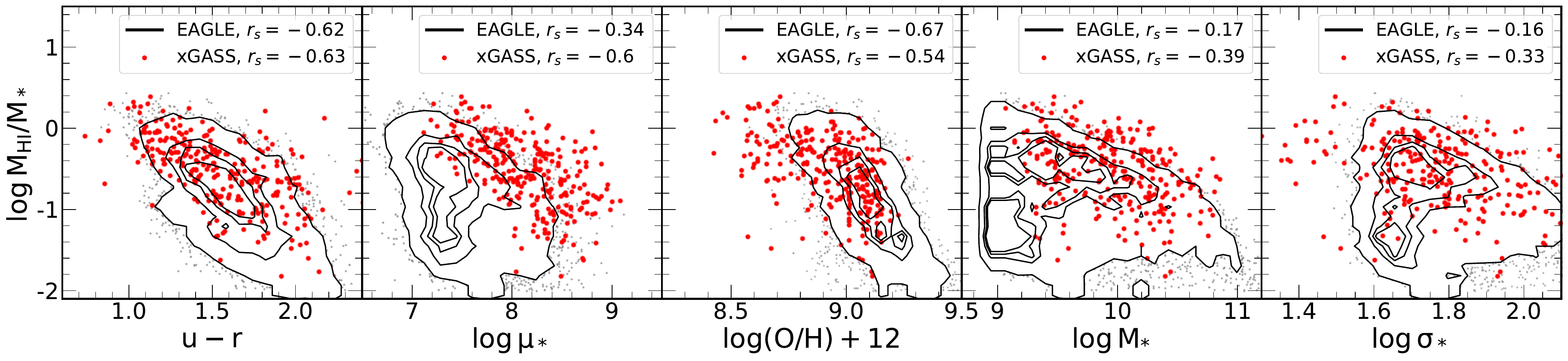}
     \caption{\hi\ fraction as a function of galaxy properties. Contours represent the simulation sample. The red dots represent the xGASS sample. The upper panel shows the TNG sample. The bottom panel shows the EAGLE sample. Spearman correlation coefficient of each sample is shown in the legend.}
     \label{fig:fHI_feature_diagram}
\end{figure*}

\autoref{fig:I_xgass} shows that neither of the two simulations can fully reproduce the ranking of feature importance in the xGASS sample. The top-ranked property in both simulations, \logoh, is ranked only in the middle in the real sample. TNG is particularly in contrast 
with the observational data: the top-ranked property $u-r$ in xGASS is now ranked at the bottom. The second-ranked property $\mu_\ast$ 
in xGASS is now ranked the second in TNG, a result that is largely (if not fully) produced by
the importance of black hole accretion rate ($\dot{M}_{\rm BH}$) found above. Compared to TNG, EAGLE appears to 
perform better in terms of $u-r$ being ranked second. The low importance of $\mu_\ast$, \mstar\ and $\sigma_\ast$ 
in EAGLE is consistent with the low importance of $\dot{M}_{\rm BH}$ found above. The gas-phase metallicity remains
important (ranks the top) in both simulations, in contrast to the observational data where it ranks the third.   

In \autoref{fig:fHI_feature_diagram} we show the \hi\ fraction as functions of the five properties in both xGASS (red dots) and the two simulations (black contours). For the simulations we use the full galaxy sample rather than the trimmed subsamples in order for better statistics. We note that the results shown in the figure remain unchange if any of the subsamples is used instead. Spearman correlation coefficients are indicated in each panel. For xGASS, the \hi\ mass fraction is clearly anti-correlated with $u-r$ and $\mu_\ast$, with high values of correlation coefficients ($r_s\sim -0.6$). A similar correlation with $u-r$ is seen in EAGLE but not in TNG 
The strong correlation with $\mu_\ast$ is seen in TNG but not in EAGLE.   
The \logoh\ shows the strongest correlation with the \hi\ mass fraction in both simulations.
The correlation is weaker in xGASS, particularly towards the low-metallicity end.   
These results are well consistent with the rankings of feature importance as shown in the 
previous figure. 

As discussed in \autoref{subsec:discussion_interpretation}, the different feature importance to the \hi\ fraction 
as found in the two simulations can be understood in terms of the different feedback processes implemented 
in the simulations. The fact that the correlation between the \hi\ fraction and the $u-r$ color 
(a good indicator of the sSFR) predicted by EAGLE is similar to that in xGASS implies that stellar feedback 
plays a more important role than AGN feedback for the relatively low-mass galaxies studied here.
The importance of $\mu_*$ in xGASS indicates that feedback processes affecting the cold gas content
of a galaxy are correlated with the structure of the galaxy in the central region. As shown earlier, 
such a correlation is generated in the TNG by the strong dependence of ${\dot M}_{\rm BH}$ 
(which determines the strength of AGN feedback) on $\mu_*$. The role of $\mu_*$ 
in EAGLE is much weaker than that in the observation. In principle, more intensive 
star formation, such as starburst, is expected to be associated with the formation of higher $\mu_*$
and with stronger feedback, and so some correlation between the HI mass fraction and $\mu_*$ 
is expected. The discrepancy thus indicates that either EAGLE underestimates such a correlation, 
or AGN feedback plays a more important role than that assumed in EAGLE, or both. 
Unfortunately, feedback effects from AGN and starburst are difficult to distinguish 
by their results for the cold gas content of galaxies. Observational constraints on 
black hole mass and AGN activities in low-mass galaxies are needed to break the 
degeneracy. Finally, the high importance of gas-phase metallicity in both simulations is not 
seen in xGASS. As discussed above, a tight correlation between the gas-phase
metallicity and the gas fraction is expected when metals generated are 
well mixed with the ISM so that the loss of metal mass in outflows is proportional to that of gas mass.
The weaker correlation seen in xGASS thus suggests that the metal and gas in real galaxies are 
not as tightly coupled as in the simulations.

\section{Summary} \label{sec:summary}

In this work, we attempt to understand the driving mechanisms for the \hi\ gas content of star-forming central galaxies in the local Universe, by examining two of the current hydrodynamical simulations, IllustrisTNG and EAGLE, as well as the xGASS galaxy sample. Applying the random forest algorithm to a variety of galaxy properties of both simulated and observed galaxies, we obtain the feature importance of the properties to the \hi-to-stellar mass ratio  ($M_{\text{H{\sc i}}}/M_\ast$). In addition, we examine the correlations of $M_{\text{H{\sc i}}}/M_\ast$ with the galaxy properties, in order to better understand the feature importance, and we compare the results between the two simulations, and between the simulations and the xGASS sample. 

Our conclusions are summarized as follows. 

\begin{itemize}
    
    \item 
The two simulations behave differently in the random forest analysis. For EAGLE, gas-phase metallicity (\logoh) and specific star formation rate (sSFR) are identified as the most important features, and both properties show a tight and negative correlation with the \hi\ mass fraction. For TNG, the two top-ranked properties are black hole accretion rate ($\dot{M}_{\rm BH}$) and \logoh, which are also negatively correlated with $M_{\text{H{\sc i}}}/M_\ast$. The differences between the two simulations can be understood from the different feedback processes adopted. In EAGLE, the \hi\ content of central galaxies is mainly regulated by stellar feedback, which drives strong gas outflows out to the virial radius of the host halo, thus efficiently suppressing gas inflow and star formation. Consequently, the central supermassive black hole cannot grow efficiently due to the limited gas supply, and AGN feedback is weak and inefficient. In TNG, in contrast, stellar feedback is too weak to efficiently suppress gas inflow, and the black hole is able to grow efficiently and heat the surrounding gas continuously through thermal mode AGN feedback. As a result, in the TNG sample, the regulation of \hi\ content is dominated by the thermal mode AGN feedback.

\item 
Neither simulation can fully reproduce the feature importance of the real galaxies in xGASS, for which the color index of $u-r$ and surface stellar mass density ($\mu_\ast$) are ranked in the first two places. EAGLE performs better than TNG in the sense that $u-r$ is ranked highly in EAGLE but at the bottom in TNG. This result implies that stellar feedback is likely to play a more dominant role than AGN feedback in driving the \hi\ content of central galaxies at low redshift. Unlike in the simulations, gas-phase metallicity is ranked only mildly for xGASS, suggesting that metals and gas caused by feedback effects in real galaxies is not as tightly coupled as in the simulations.
\end{itemize}

Our results demonstrate that observations of \hi\ content can provide strong constraints on gas-related subgrid models in cosmological hydrodynamic simulations, which are able to reproduce both stellar and cold gas content of low-redshift galaxies but using different recipes. In the next decade spatially resolved \hi\ observations will be available for large samples of galaxies and provide more constraints for the models, thanks to the many new-generation \hi\ surveys (e.g., WALLABY, \citealt{WALLABY}; MIGHTEE-HI, \citealt{MIGHTEE-HI}; LADUMA \citealt{LADUMA}). 

\begin{acknowledgments}

We thank Robert A. Crain for kindly sharing the \hi\ data of EAGLE.
This work is supported by the National Key R\&D Program of China (grant NO. 2022YFA1602902), and the National Natural Science Foundation of China (grant Nos. 12433003, 11821303, 11973030).
This work has made use of the following software: Numpy \citep{Numpy}, Scipy \citep{SciPy}, Scikit-learn \citep{scikit-learn}, Matplotlib \citep{Matplotlib}, emcee \citep{emcee}, Astropy \citep{Astropy2013,Astropy2018,Astropy2022}, and h5py \citep{H5py}.

Funding for the SDSS and SDSS-II has been provided by the Alfred P. Sloan Foundation, the Participating Institutions, the National Science Foundation, the U.S. Department of Energy, the National Aeronautics and Space Administration, the Japanese Monbukagakusho, the Max Planck Society, and the Higher Education Funding Council for England. The SDSS Web Site is http://www.sdss.org/.

The SDSS is managed by the Astrophysical Research Consortium for the Participating Institutions. The Participating Institutions are the American Museum of Natural History, Astrophysical Institute Potsdam, University of Basel, University of Cambridge, Case Western Reserve University, University of Chicago, Drexel University, Fermilab, the Institute for Advanced Study, the Japan Participation Group, Johns Hopkins University, the Joint Institute for Nuclear Astrophysics, the Kavli Institute for Particle Astrophysics and Cosmology, the Korean Scientist Group, the Chinese Academy of Sciences (LAMOST), Los Alamos National Laboratory, the Max-Planck-Institute for Astronomy (MPIA), the Max-Planck-Institute for Astrophysics (MPA), New Mexico State University, Ohio State University, University of Pittsburgh, University of Portsmouth, Princeton University, the United States Naval Observatory, and the University of Washington.

\end{acknowledgments}

\bibliography{refs}{}
\bibliographystyle{aasjournal}

\end{document}